\documentstyle[11pt]{article}

\begin{document}

\title{  Time reversal odd fragmentation functions in semi-inclusive \\
scattering of polarized leptons from unpolarized hadrons}

\author{
J. Levelt \\
\mbox{} \\
Institut f\"ur Theoretische Physik, University of Erlangen-N\"urnberg, \\
Staudtstrasse 7, D-91508 Erlangen, Germany \\
\mbox{}
\\ and \\
\mbox{} \\
 P.J. Mulders\thanks{
\  also at Department of Physics and Astronomy, Free University, Amsterdam}
\\
\mbox{ } \\
National Institute for Nuclear Physics and High Energy Physics \\
 (NIKHEF-K), P.O. Box 41882, NL-1009 DB Amsterdam, the Netherlands}
\date{}
 \maketitle

\begin{abstract}
Semi-inclusive deep inelastic scattering of polarized leptons off hadrons
enables one to measure the antisymmetric part of the hadron tensor.
For unpolarized hadrons this piece is odd under
time reversal. In deep inelastic scattering it shows up as a $\langle
\sin \phi \rangle$ asymmetry for the produced hadrons. This asymmetry can
be expressed as the product of a twist-three
"hadron $\rightarrow$ quark" distribution function
and a time reversal odd twist-two "quark $\rightarrow$ hadron"
fragmentation function.
This fragmentation function can only be measured for nonzero transverse
momenta of the produced hadron.

\end{abstract}

\vspace{5.5 cm}

\noindent
NIKHEF 94-P5\\
hep-ph/9408257 \\
August 94\\
submitted to Physics Letters B
\newpage

In this paper we study deep inelastic leptoproduction of hadrons. To be
precise we investigate the process
\begin{equation}
\ell + H \ \longrightarrow \ \ell^\prime + h + X,
\end{equation}
where $\ell$ and $\ell^\prime$ are the incoming and scattered lepton with
momenta $k$ and $k^\prime$, $H$ is the hadronic target with mass $M$ and
momentum $P$, $h$ is the produced hadron with mass $M_h$ and momentum
$P_h$ and $X$ is the unobserved rest of the final state. The differential
cross section for this process is written as a product of a leptonic and
hadronic tensor,
\begin{equation}
\frac{d\sigma}{dx_{\scriptscriptstyle B} \,dz\,dy\, d^2\mbox{\boldmath $q$}_T}
= \frac{\pi\,\alpha^2\,y\,z}{2\,Q^4}
L_{\mu \nu}\, 2M{\cal W}^{\mu \nu},
\end{equation}
where we have used the invariants expressed in the external momenta and the
momentum $q$ of the exchanged virtual photon (we will limit ourselves to
electromagnetic interactions), which is spacelike and large ($-q^2$ = $Q^2 \gg
M^2$). The invariants used are\footnote{
With an approximate equal sign we indicate relations valid up to order
$1/Q^2$.}
\begin{equation}
x_{\scriptscriptstyle B} = \frac{Q^2}{2\,P\cdot q},\quad
z = \frac{P\cdot P_h}{P\cdot q} \approx -\frac{2\,P_h\cdot q}{Q^2},\quad
y = \frac{P\cdot q}{P\cdot k}.
\label{scalingvariables}
\end{equation}
The (two-component) vector $\mbox{\boldmath $q$}_T$ is the transverse momentum
of the photon in the frame in which the  hadrons $H$ and $h$ are collinear.
This translates to a perpendicular momentum of the produced hadron equal to
${{\mbox{ \boldmath $P$}}}_{h\perp}$ = $-z\mbox{\boldmath $q$}_T$ in
the frame in which the target hadron $H$ and the photon are collinear. The
leptonic tensor for polarized leptons with helicity $\lambda = \pm 1$ is of the
form
\begin{equation} \label{leptten}
L_{\mu\nu} (k, k^\prime, \lambda)
= \left(2 k_{\mu}k^\prime_{\nu} + 2 k_{\nu}k^\prime_{\mu}
- Q^2 g_{\mu\nu} +2i\lambda \epsilon_{\mu\nu\rho\sigma}
q^\rho k^{\sigma} \right) .
\end{equation}
and the (unpolarized) hadronic tensor is of the form
\begin{eqnarray}
2M{\cal W}_{\mu\nu}( q, P, P_h)
& = & \frac{1}{(2\pi)^4}
\int \frac{d^3 P_X}{(2\pi)^3 2P_X^0}\,
 (2\pi)^4 \delta^4 (q + P - P_X - P_h)
\nonumber \\
& & \qquad \qquad \times \langle P |J_\mu (0)|P_X, P_h \rangle
\langle P_X, P_h |J_\nu (0)|P \rangle ,
\label{hadrten}
\end{eqnarray}
where averaging over initial state spins and summation over final state
spins is assumed.

In all of our considerations we will neglect all contributions of order
$1/Q^2$, but keep all contributions of order $1/Q$, as it turns out
that the time reversal odd fragmentation function, that we investigate,
contributes at the ${\cal O}(1/Q)$ level.
We will also not discuss the factorization of the process
and the inclusion of radiative corrections. We assume that we can use the
diagrammatic expansion for the hard process as outlined by Ellis, Furm\'anski
and Petronzio (EFP) \cite{EFP}. In this approach diagrams are separated in
a hard scattering part and soft correlation functions \cite{Collinsbook}.

The diagrammatic expansion of EFP is useful for deep inelastic scattering,
because it in an appropriately chosen (color) gauge corresponds to an expansion
in powers of $Q^{-1}$. This twist expansion as used here is discussed in
refs \cite{Jaffe83,JaffeJi,LM,TM}. The Born diagrams in this expansion are
given in Fig. 1 and give \cite{LM}
\begin{equation}
2M\,{\cal W}_{\mu \nu}(q,P,P_h) =
\frac{1}{2}\,\int d^4p\,d^4k\,\delta^4(p+q-k)\,
\mbox{Tr}\left(\gamma_\mu \Phi(p) \gamma_\nu \Delta(k)\right)
+ \left\{ \begin{array}{c} q \leftrightarrow -q \\ \mu \leftrightarrow \nu
\end{array} \right\},
\label{born}
\end{equation}
where the (soft) quark correlation functions \cite{Soper,Jaffe83,CS77} are
given by
\begin{eqnarray}
& & \Phi_{ij}(p) = \frac{1}{(2\pi)^4} \int d^4 x \, e^{ip\cdot x}\,
\langle P \vert \overline \psi_j (0) \psi_i(x) \vert P \rangle, \label{qq1}
\\
& & \Delta_{ij}(k) = \sum_X \frac{1}{(2\pi)^4} \int d^4x\, e^{ik\cdot x}\,
\langle 0 \vert \psi_i(x) \vert P_h,X \rangle \langle P_h,X \vert \overline
\psi_j(0) \vert 0 \rangle. \label{qq2}
\end{eqnarray}
In order to analyze the result it is useful to use lightcone coordinates for
the external momenta $P$, $P_h$ and $q$ and the integration variables $p$ and
$k$. Using the representation $p$ = $[p^-,p^+,\mbox{\boldmath $p$}_T]$ where
$p^\pm = (p^0 \pm p^3)/\sqrt{2}$ we write in a frame where $H$ and $h$ are
collinear
\begin{eqnarray}
P & = & \left[ \frac{x_{\scriptscriptstyle B} M^2}{A\sqrt{2}},
\frac{A}{x_{\scriptscriptstyle B} \sqrt{2}}, \mbox{$\bf 0$}_{T} \right],
\nonumber \\
p & = & \left[ p^-, \frac{A}{\sqrt{2}}, \mbox{\boldmath $p$}_T \right],
\nonumber \\
q & = & \left[ \frac{Q^2}{A\sqrt{2}}, -\frac{A}{\sqrt{2}},
\mbox{\boldmath $q$}_T \right], \\
k & = & \left[ \frac{Q^2}{A\sqrt{2}}, k^+, \mbox{\boldmath $k$}_T \right],
\nonumber \\
P_h & = & \left[ \frac{z Q^2}{A\sqrt{2}}, -\frac{A M_h^2}{z Q^2 \sqrt{2}},
\mbox{$\bf 0$}_{T} \right],
\nonumber
\end{eqnarray}
This parametrization is consistent with $p+q = k$ and has been made under the
assumption that the quark momenta are limited to a hadronic scale
of the order of a few hundred MeV. More precisely, $p^2$, $P\cdot p$
and $k^2$, $P_h\cdot k$ are of hadronic scale.

In the deep inelastic limit the process factorizes. For the $H \rightarrow q$
part one can integrate over $p^-$. One can then analyze the Dirac content
of the projections \cite{RS79}
\begin{equation}
\Phi^{[\Gamma]}(x_{\scriptscriptstyle B},\mbox{\boldmath $p$}_T)
= \left. \frac{1}{2} \int dp^-\ Tr(\Phi\,\Gamma )
\right|_{p^+ = x_{\scriptscriptstyle B} P^+,\ \mbox{\boldmath $p$}_T}.
\end{equation}
Using for $\Phi$ the constraints imposed by hermiticity,
parity and time reversal invariance one obtains for an unpolarized hadron
\begin{eqnarray}
& & \Phi^{[\gamma^+]} = f_1(x_{\scriptscriptstyle B} ,\mbox{\boldmath $p$}_T),
\label{Phi1} \\
& & \Phi^{[1]} = \frac{M}{P^+}\,e(x_{\scriptscriptstyle B} ,
\mbox{\boldmath $p$}_T), \label{Phi2} \\
& & \Phi^{[\gamma^i]} = \frac{p_T^i}{P^+}\,f^\perp(x_{\scriptscriptstyle B} ,
\mbox{\boldmath $p$}_T), \label{Phi3}
\end{eqnarray}
where $f_1$ is the well-known twist-two quark distribution function, in
which we have kept the $\mbox{\boldmath $p$}_T$-dependence and $e$ and
$f^\perp$ are the twist-three profile functions discussed in
ref.~\cite{JaffeJi} and \cite{LM}, respectively.
The twist of the profile functions is obtained from the power
$1/(P^+)^{t-2}$ appearing in the correlation function.
Note that only after integrating over transverse momenta and expansion in terms
of local operators a comparison with the twist known from the operator product
expansion can be made.

Similarly, for the $q \rightarrow h$ part one can integrate over $k^+$ and
analyze the Dirac content of the projections
\begin{equation}
\Delta^{[\Gamma]}(z,\mbox{\boldmath $k$}_T)  =
\left. \frac{1}{4z}\int dk^+\ Tr(\Delta\,\Gamma)\right|_{k^- =
P_h^-/z,\ \mbox{\boldmath $k$}_T}.
\end{equation}
As the definition of $\Delta$ involves states $\vert P_h,X\rangle$, which are
out-states, one cannot use constraints from time reversal invariance. As a
consequence one obtains as the most general Dirac content
\begin{eqnarray}
& & \Delta^{[\gamma^-]} =
D_1(z,-z\mbox{\boldmath $k$}_T), \label{Delta1}
\\ & & \Delta^{[i \sigma^{i-} \gamma_5]} =
\frac{\epsilon_T^{ij} k_{T j}}{M_h}\,H_1^\perp(z,-z\mbox{\boldmath $k$}_T),
\label{Delta2} \\
& & \Delta^{[1]} =
\frac{M_h}{P_h^-}\,E(z,-z\mbox{\boldmath $k$}_T) , \label{Delta3}
\\ & & \Delta^{[\gamma^i]} =
\frac{k_T^i}{P_h^-}\,D^\perp(z,-z\mbox{\boldmath $k$}_T), \label{Delta4}
\\ & & \Delta^{[ i \sigma^{ij} \gamma_5]} =
\frac{M_h\,\epsilon^{ij}}{P_h^-}\,H(z,-z\mbox{\boldmath $k$}_T),
\label{Delta5}
\end{eqnarray}
where $D_1$ is the well-known twist-two quark fragmentation function with
the $\mbox{\boldmath $k$}_T$-dependence kept. This dependence translates
into a dependence on the transverse momenta of produced hadrons.
Furthermore a time reversal odd twist-two  fragmentation function
$H_{1}^\perp$ appears, which can only be measured when the dependence
on the transverse momenta is measured, e.g. in azimuthal asymmetries.
This function $H_1^\perp$ has been discussed in ref.~\cite{Co}
for polarized leptoproduction as a means to probe transversely polarized
quarks. At the twist-three level one finds the profile functions
$E$ (ref.~\cite{JaffeJi}), $D^\perp$ (ref.~\cite{LM}) and another
time reversal odd function $H$.

One can gain some insight in the nature of these time reversal odd functions
by considering the case that $hX$ would have only one possible channel.
Using $\vert P_h,X;out\rangle$ =  $e^{2i\delta_{hX}}\,\vert
P_h,X;in\rangle$ (see \cite{gas}), the constraints from time-reversal then
would imply that the functions $D_1$, $E$ and $D^\perp$ are proportional
to $\cos \delta_{hX}$, while the functions $H_1^\perp$ and $H$ are
proportional to $\sin \delta_{hX}$.

The twist-two functions are the ones that appear if one evaluates the result
for the hadronic tensor in leading order. This yields
\begin{equation}
2M\,{\cal W}_{\mu \nu}
= \left( \frac{\tilde P_\mu \tilde P_\nu}{\tilde P^2} \right) \,
2z\,I[f_1\,D_1],
\end{equation}
where $\tilde P$ = $P - (P\cdot q/q^2)\,q$, and $\tilde P^2$ = $Q^2/4x^2$,
while the convolution $I[f_1\,D_1]$ is given by
\begin{equation}
I[f_1 \,D_1] =
\int d^2p_\perp\,d^2k_\perp\, \delta^2(\mbox{\boldmath $p$}_\perp +
\mbox{\boldmath $q$}_T - \mbox{\boldmath $k$}_\perp)\,
f_1(x_{\scriptscriptstyle B},\mbox{\boldmath $p$}_\perp)\,
D_1(z,-z\mbox{\boldmath $k$}_\perp). \label{conv}
\end{equation}
As illustration we will consider the situation in which
\begin{eqnarray}
&& f(x_{\scriptscriptstyle B},\mbox{\boldmath $p$}_\perp) =
f(x_{\scriptscriptstyle B},0)\,\exp (- R_H^2 \mbox{\boldmath $p$}_\perp^2),
\label{f1app} \\
&& D(z,-z\mbox{\boldmath $k$}_\perp^2) = D(z,0) \,
\exp (- R_h^2 \mbox{\boldmath $k$}_\perp^2). \label{D1app}
\end{eqnarray}
In that case the above integral can be written as
\begin{equation}
I[f_1 \,D_1] =
\frac{\pi}{R_H^2 + R_h^2}\,\exp \left\lgroup - \frac{\mbox{\boldmath $q$}_T^2
R_H^2 R_h^2}{R_H^2 + R_h^2} \right\rgroup \,f_1(x_{\scriptscriptstyle B},
\mbox{$\bf 0$}_{T})\,D_1(z,\mbox{$\bf 0$}_{T}).
\end{equation}
Integrating over the perpendicular momenta of the produced hadron, i.e. over
$\mbox{\boldmath $q$}_T$ one obtains the familiar factorized form
$f_1(x_{\scriptscriptstyle B})\, D_1(z)$, where
$f_1(x_{\scriptscriptstyle B})$ =
$\int d^2p_T\,f_1(x_{\scriptscriptstyle B},\mbox{\boldmath $p$}_T)$ is the
lightcone momentum distribution and $D_1(z)$ = $\int d^2P_{h\perp}
D_1(z,{{\mbox{ \boldmath $P$}}}_{h\perp})$ is the
fragmentation function. Finally it should be noted that everywhere the
summation over quark flavors has been suppressed. Together with the antiquark
part, it can easily be reinstated by adding a summation over flavors of quarks
{\em and} antiquarks weighted with the charge squared ($e_i^2$) and labeling
the functions as $f_{1}^{i/H}$ and $D_1^{h/i}$ respectively.

The Born contribution to ${\cal W}_{\mu \nu}$ also gives a contribution at
subleading order (proportional to $1/Q$). These involve the twist-three
functions defined above. At the same order, however, also contributions
from diagrams with gluons, as shown in Fig. 2 for quarks, must be included.
They give in addition to the result in Eq.~\ref{born} contributions that
involve quark - quark - gluon correlation functions. It turns out that they
in the gauges $A^+ = 0$ (for $H\rightarrow q$ part) and $A^- = 0$ (for
$q\rightarrow h$ part) at the twist-three level contribute through
the bilocal matrix elements
\begin{eqnarray}
& & F^\alpha_{ij}(p) = \frac{1}{(2\pi)^4}\int d^4x\ e^{i\,p\cdot x}
\langle P \vert \overline \psi_j(0) A_T^\alpha (x) \psi_i(x)
\vert P \rangle. \label{qqG1}\\
& & M^\alpha_{ij}(k) = \frac{1}{(2\pi)^4}\sum_X \int d^4x\ e^{i\,k\cdot
x} \langle 0 \vert A_T^\alpha (x) \psi_i(x)\vert P_h,X \rangle \langle P_h, X
\vert \overline \psi_j(0) \vert 0 \rangle. \label{qqG2}
\end{eqnarray}
Using the equations of motion for the quark fields, these
can be reexpressed in the quark - quark correlation functions $\Phi$ and
$\Delta$.

At this point it is appropriate to mention that in order to have a color gauge
invariant definition one must include a color link in the bilocal
quark-quark correlation functions in Eqs~\ref{qq1} and \ref{qq2} and similarly
in the quark-quark-gluon correlation functions in Eq.~\ref{qqG1} and
\ref{qqG2}. For correlation function integrated over $k^-$ {\em and}
$\mbox{\boldmath $k$}_T$ one is only sensitive to the nonlocality in the $x^-$
direction. In that case a link $L(0,x) = {\cal P}\,\exp [i g \int_0^x d s\cdot
A(s)]$ with a straight path can be inserted in the definition of profile
functions. Such a link operator becomes unity in the gauge $A^+ = 0$. When
transverse momentum is observed and one only integrates over $k^-$,
one also becomes sensitive to separation in the transverse direction, although
still $x^+$ = 0. In that case one also needs to fix the gauge freedom
affecting $A_T$. This requires a choice of boundary conditions \cite{KS70} for
$A_T$ and correspondingly a particular choice of path in the link operator.
Although it is not hard to find some way to do this in such a way that the link
operator reduces to unity after fixing the gauge, it remains to be proven
that this can be combined with a proof for factorization at the twist three
level.

The full (electromagnetically gauge invariant) result of the diagrammatic
expansion up to ${\cal O}(1/Q)$ reads
\begin{eqnarray}
2M\,{\cal W}_{\mu \nu} & = &
\int d^2p_T\,d^2k_T\, \delta^2(\mbox{\boldmath $p$}_T + \mbox{\boldmath $q$}_T
- \mbox{\boldmath $k$}_T)\, \nonumber \\
& & \quad \times \Biggl[ \left( \frac{q_\mu q_\nu}{q^2} - g_{\mu \nu}
+ \frac{\tilde P_\mu \tilde P_\nu}{\tilde P^2} \right) \,
2z\,f_1\,D_1 \nonumber \\
& & \qquad + \frac{(\tilde P_\mu p_{\perp \nu} + \tilde P_\nu p_{\perp \mu})}
{Q^2}\,8x_{\scriptscriptstyle B} z\,f^\perp\, D_1 \nonumber \\
& & \qquad + \frac{(\tilde P_\mu k_{\perp^\nu} + \tilde P_\nu k_{\perp \mu})}
{Q^2}\,8 z\,f_1\, \left(\frac{1}{z}\,D^\perp - D_1 \right)
\nonumber \\
& & \qquad + i\,\frac{(\tilde P_\mu k_{\perp \nu} - \tilde P_\nu
k_{\perp \mu})} {Q^2}\,8 z\,\frac{M}{M_h}\,\left(x_{\scriptscriptstyle B} e
- \frac{m}{M}\,f_1 \right)\,H_1^\perp
\Biggr],
\end{eqnarray}
where $p_\perp$ is constructed from the vector that in the frame where the
hadrons ($H$ and $h$) are collinear is $p_T$ = $[0,0,\mbox{\boldmath $p$}_T]$
by subtracting the projection along $q$, i.e. $p_\perp$ = $p_T$ +
$(\mbox{\boldmath $p$}_T\cdot \mbox{\boldmath $q$}_T/q^2)\,q$.
This vector is orthogonal to $q$. Note, however that neglecting $1/Q^2$
contributions, the perpendicular components are the same, i.e.
$\mbox{\boldmath $p$}_T$ = $\mbox{\boldmath $p$}_\perp$.

In order to investigate the observable consequences of the various terms in
the hadronic tensor, one needs to consider the contraction with the leptonic
tensor. This gives
\begin{eqnarray}
& & L^{\mu \nu} \left( \frac{q_\mu q_\nu}{q^2} - g_{\mu \nu}
+ \frac{\tilde P_\mu \tilde P_\nu}{\tilde P^2} \right)
= \frac{4 Q^2}{y^2}\,\left( \frac{y^2}{2} + 1 - y \right), \\
& & L^{\mu \nu} \,
\left( \frac{\tilde P_\mu p_{\perp \nu} + \tilde P_\nu p_{\perp \mu}}
{Q^2}\right)
= -\frac{4 Q^2}{y^2}\,\frac{\vert \mbox{\boldmath $p$}_T \vert}{Q}\,
\frac{(2-y)\sqrt{1-y}}{2x}\, \cos \phi_p, \\
& & L^{\mu \nu} \,
\left( i\frac{\tilde P_\mu k_{\perp \nu} - \tilde P_\nu k_{\perp \mu}}
{Q^2}\right)
= -\frac{4 Q^2}{y^2}\,\frac{\vert \mbox{\boldmath $k$}_T \vert}{Q}\,
\frac{y\sqrt{1-y}}{2x}\, \lambda \sin \phi_k,
\end{eqnarray}
where the azimuthal angles are those between the transverse vectors and
the (lepton) scattering plane, e.g. $\mbox{\boldmath $p$}_T\cdot
\mbox{\boldmath $q$}_T$ = $\vert \mbox{\boldmath $p$}_T \vert \,
\vert \mbox{\boldmath $q$}_T \vert\,\cos \phi_p$.

As the most general result one can consider the case where one measures the
perpendicular direction of the struck quark from the jet direction
($\mbox{\boldmath $p$}_\perp$)
and the momentum of one specific hadron ($h$) belonging to this jet,
characterized by the lightcone momentum fraction $z$ and the perpendicular
momentum $\mbox{\boldmath $p$}_{h\perp}$. One obtains (using
$\mbox{\boldmath $k$}_\perp$ = $\mbox{\boldmath $p$}_\perp -
\mbox{boldmath $P$}_{h\perp}/z$)
\begin{eqnarray}
	& & \frac{d\sigma^{(\ell+H\rightarrow \ell^\prime
         + h + \mbox{\scriptsize jet} + X)}}
	{ dx_{\scriptscriptstyle B} dydzd^{\,2}\mbox{\boldmath $P$}_{h\perp}
	d^{\, 2} \mbox{\boldmath $p$}_{\perp}} =
        \frac{4\pi\alpha^2\,s}{Q^4}\,
	 \Bigg\{\left\lgroup
	\frac{y^2}{2}+1-y\right\rgroup x_{\scriptscriptstyle B} f_1\, D_{1}
	\nonumber \\ &&\qquad\qquad
        \mbox{}+2(2-y)\sqrt{1-y}\,\cos\phi_h\,\frac{|\mbox{\boldmath
	$P$}_{h\perp}|}{zQ}
	\Bigg[ x_{\scriptscriptstyle B} f_1\, \left\lgroup\frac{1}{z}
	D^{\perp} -D_1 \right\rgroup \Bigg] \nonumber \\
	&&\qquad\qquad \mbox{}+2(2-y)\sqrt{1-y}\,\cos\phi_j\,
	\frac{|\mbox{\boldmath $p$}_{\perp}|}{Q}
	\Bigg[ 	- \frac{x_{\scriptscriptstyle B}}{z}\,f_1 \, D^{\perp}
	+x_{\scriptscriptstyle B}
        \left\lgroup f_1 -x_{\scriptscriptstyle B} f^{\perp}
        \right\rgroup D_1 \Bigg] \nonumber \\
	&&\qquad\qquad\mbox{}+2y\sqrt{1-y}\,\lambda\,\sin\phi_h\,
        \frac{|\mbox{\boldmath $P$}_{h\perp}|}{zQ}
	\Bigg[ \frac{Mx_{\scriptscriptstyle B}}{M_h}
        \left(x_{\scriptscriptstyle B} e
        - \frac{m}{M}\, f_1\right) H_1^{\perp} \Bigg]
        \nonumber \\
	&&\qquad\qquad\mbox{}-2y\sqrt{1-y}\,\lambda\,\sin\phi_j\,
	\frac{|\mbox{\boldmath $p$}_{\perp}|}{Q}
	\Bigg[ \frac{Mx_{\scriptscriptstyle B}}{M_h}
        \left(x_{\scriptscriptstyle B} e
        - \frac{m}{M}\, f_1\right) H_1^{\perp} \Bigg]
        \Bigg\},  \label{cross}
\end{eqnarray}
where the $H \rightarrow q$ profile functions ($f_1$, $f^\perp$ and $e$) depend
on $x_{\scriptscriptstyle B}$ and $\mbox{\boldmath $p$}_\perp^2$, while the
$q \rightarrow h$ profile functions ($D_1$, $D^\perp$ and $H_1^\perp$) depend
on $z$ and $(\mbox{\boldmath $P$}_{h\perp} - z\mbox{\boldmath $p$}_\perp)^2$.

The theoretically simplest results are the ones obtained by integrating over
the momenta of produced hadrons \cite{LM}. Integrating over the perpendicular
momenta one has
\begin{eqnarray}
	\frac{d\sigma^{(\ell+H\rightarrow \ell^\prime
         + h + \mbox{\scriptsize jet} + X)}}
	{ dx_{\scriptscriptstyle B} dydz
	d^{\, 2} \mbox{\boldmath $p$}_{\perp}} & = &
        \frac{4\pi\alpha^2\,s}{Q^4}\, \Bigg\{\left\lgroup
	\frac{y^2}{2}+1-y\right\rgroup x_{\scriptscriptstyle B}
        f_1(x_{\scriptscriptstyle B},
        \mbox{\boldmath $p$}_{\perp})\, D_{1}(z)
	 \nonumber \\
	&&\mbox{}-2(2-y)\sqrt{1-y}\,\cos\phi_j\,
	\frac{|\mbox{\boldmath $p$}_{\perp}|}{Q}\,
	x^2_{\scriptscriptstyle B} f^{\perp}(x_{\scriptscriptstyle B},
        \mbox{\boldmath $p$}_{\perp})
        \,D_1(z) \Bigg\}.
\end{eqnarray}
This shows that the ratio $x_{\scriptscriptstyle B}
f^\perp(x_{\scriptscriptstyle B},\mbox{\boldmath $p$}_{\perp})/
f_1(x_{\scriptscriptstyle B},\mbox{\boldmath $p$}_{\perp})$ can be obtained
most cleanly from the azimuthal $\langle \cos \phi_j \rangle$  asymmetry of
the produced jet. Note that this ratio is an extension of the naive parton
result of unity \cite{cahn}. A calculation in the bag model is shown in
ref.~\cite{LM2}.

The measurement of the time reversal odd fragmentation function $H_1^\perp$
is possible from the general cross section given above. However, the detection
of a jet in the forward direction will be hard in view of the fact that
one is interested in a twist-three piece ($\propto 1/Q$). This makes it more
realistic to consider $\ell + H \rightarrow \ell^\prime +
h + X$. Only convolutions of
profile functions from the distribution and fragmentation part will appear
in this case. We will consider again the approximations for the transverse
momentum dependence as in Eqs.~\ref{f1app} and \ref{D1app} and express the
results in convolutions $I$ as defined in Eq.~\ref{conv}.
The result for the cross section for electroproduction of a hadron $h$ up
to ${\cal O}(1/Q)$ becomes
\begin{eqnarray}
	&&\frac{d\sigma^{(\ell+H\rightarrow \ell^\prime + h + X)}}
	{ dx_{\scriptscriptstyle B} dydz
	d^{\, 2} \mbox{\boldmath $p$}_{\perp}} =
        \frac{4\pi\alpha^2\,s}{Q^4}\, \Bigg\{\left\lgroup
	\frac{y^2}{2}+1-y\right\rgroup x_{\scriptscriptstyle B}
        I[f_1\, D_1] \nonumber \\
        &&\qquad\mbox{}+2(2-y)\sqrt{1-y}\,\cos\phi_h\,
        \frac{|\mbox{\boldmath $P$}_{h\perp}|}{zQ}
	\left\lgroup \frac{ R_h^2\, x_{\scriptscriptstyle B}}{R_H^2 + R_h^2}\,
        I\left[ f_1 \left( D_1 - \frac{1}{z}
	D^{\perp} \right) \right]
	- \frac{ R_H^2 \,x_{\scriptscriptstyle B}^2}{R_H^2 + R_h^2}\,
        I[f^\perp\, D_1 ] \right\rgroup \nonumber \\
	&&\qquad\mbox{}+2y\sqrt{1-y}\,\lambda \sin\phi_h\,
        \frac{|\mbox{\boldmath $P$}_{h\perp}|}{zQ}
	\left\lgroup \frac{Mx_{\scriptscriptstyle B}}{M_h}
        \frac{R_h^2}{R_H^2 + R_h^2}\,
        I\left[\left(\frac{m}{M}\, f_1 - x_{\scriptscriptstyle B} e \right)
        H_1^{\perp} \right] \right\rgroup \Bigg\}.
\end{eqnarray}
The time reversal odd fragmentation function $H_1^\perp$ is a leading twist
fragmentation function which appears as a consequence of the fact that
the states $\vert p_h, X \rangle$ in the definition of $\Delta$ are out-states.
This means that it is nonzero because of the interactions of the outgoing
hadron $h$. Furthermore, this time reversal odd structure function appears
in the cross section as part of a product that also contains the profile
function $e$. This chirally odd function has been discussed in
ref.~\cite{JaffeJi} and an estimate using the bag model has been given.
We note that the combination that appears in the cross sections corresponds
to a pure quark-quark-gluon correlation function, to be precise
$i(Mx_{\scriptscriptstyle B}\,e - m\,f_1)$ = $(1/2)\int dk^-\,
\mbox{Tr}\,(g\,F_\alpha\,\sigma^{\alpha +})$.

In principle the extraction of the $\sin \phi_h$ asymmetry for pions does not
seem to be extremely difficult. One does not need target polarization, while
the result is proportional to the lepton helicity. The azimuthal structure in
the quark transverse momenta plays the key role in this. The precise $\vert$
{\boldmath $p$}$_{\perp}$ $\vert$ dependence and dilution of it by QCD
corrections (Sudakov effects \cite{long}) are less relevant. It is important
to emphasize that the {\boldmath $p$}$_{\perp}$-dependence that we have
discussed is the low-momentum range, say below 1 GeV. Here the \mbox{\boldmath
$p$}$_{\perp}$-dependence is dominated by the intrinsic transverse momentum of
the profile functions, rather than by perturbative QCD processes \cite{AM}
that dominate the high transverse momenta  \cite{E665,conrad,baker}. The fact
that one deals with a twist three structure function furthermore requires not
too large values for the momentum transfer $Q$. It is important, however, to
have good particle identification and sufficient azimuthal resolution in the
forward direction.

We acknowledge discussions with R.T. Tangerman and J. Collins. The work of
J.L. is supported by the Bundesministerium f\"ur Forschung und Technologie
and that of P.J.M. is supported by the foundation for Fundamental Research
of Matter (FOM) and the National Organization for Scientific Research (NWO).
We also thank the European Centere for Theoretical Studies in Nuclear Physics
and Related Areas (ECT*) in Trento for its hospitality and for partial support
during the final stage of this work.

\newpage

\begin{center} {\bf List of Figures   } \end{center}

\begin{description}
\item{\bf fig. 1: } The Born terms in the expansion of the hadronic tensor.

\item{\bf fig. 2: } The contributions from quark - quark - gluon correlation
functions in the expansion of the hadronic tensor. Only the quark part is
shown.

\end{description}

\end{document}